\documentclass[12pt]{iopart}

\usepackage{iopams}
  
\usepackage{graphicx}

\begin{document}

\title[Exact solutions to the Lighthill--Whitham--Richards--Payne traffic flow
equations II]{Some exact solutions to the
Lighthill--Whitham--Richards--Payne\\ traffic flow equations II:
moderate congestion}

\author{E Infeld$^1$, G Rowlands$^2$ and A A Skorupski$^1$}

\address{
$^1$ National Centre for Nuclear Research, Ho\.{z}a 69, 00--681 Warsaw, Poland\\
$^2$ Department of Physics, University of Warwick, Coventry CV4 7AL, UK
}

\eads{\mailto{einfeld@fuw.edu.pl}, \mailto{g.rowlands@warwick.ac.uk},  
\mailto{askor@fuw.edu.pl}}

\begin{abstract}
We find a further class of exact solutions to the Lighthill--Whitham--Richards%
--Payne (LWRP) traffic flow equations. As before, using two consecutive Lagrangian
transformations, a linearization is achieved. Next, depending on the initial
density, we either obtain exact formulae for the dependence of the car
density and velocity on $x,t$, or else, failing that, the same result in a
parametric representation. The calculation always involves two possible
factorizations of a consistency condition. Both must be considered. In physical
terms, the lineup usually separates into two offshoots at different velocities.
Each velocity soon becomes uniform. This outcome in many ways resembles not only
Rowlands, Infeld and Skorupski 2013 \textit{J. Phys. A: Math. Theor.} \textbf{46}
365202 (part I) but also the two soliton solution to the Korteweg--de Vries
equation. This paper can be read independently of part I. This explains unavoidable
repetitions. Possible uses of both papers in checking numerical codes are indicated
at the end. Since LWRP, numerous more elaborate models, including multiple lanes,
traffic jams, tollgates etc. abound in the literature. However, we present an exact
solution. These are few and far between, other then found by inverse scattering.
The literature for various models, including ours, is given. The methods used here
and in part I may be useful in solving other problems, such as shallow water flow. 
\end{abstract}


\submitto{\JPA}

\section{General history. Formulation of the model}

Some nonlinear, partial differential equations
of physics, not integrable by inverse scattering or else an
inversion of variables, yield their secrets to Lagrangian coordinate methods
\cite{InfRol1}--\cite{Jord}.
Here we will treat one such equation pair and see a combination of two
Lagrangian transformations enable us to solve the one lane moderately congested
traffic flow problem explicitly. A further class of solutions is found in parametric
form. Calculations augment and reinforce those of part I \cite{partI}.

In 1955, James Lighthill and Gerald Whitham formulated an
equation describing single lane traffic flow, assumed congested enough to justify
a fluid model \cite{Light}.
Richards published in the following year \cite{Rich}. Next Payne
\cite{Payne} and Whitham \cite{Whith1} added a second equation and replaced the LWR
equation with standard continuity. We will call this pair LWRP. Recently the
literature on both models has grown considerably, see for example the books by Kern
\cite{Kern1} and further references \cite{Chandl}--\cite{Treiber}.

Extensions to more than one lane, lane changing, discrete models, higher order
effects, as well as numerical work, prevail. One of the original authors has found a
Toda lattice like solution to the discrete version of Newell \cite{New}, see
\cite{Whith2}. In a future paper, we will see if the methods introduced here can be
applied to some of these recent extensions of LWRP and LWR.

Other models have been developed \cite{Treiber}, and \cite{Rascle}--\cite{Dolfin}.
However, here we will concentrate on LWRP, remembering that any progress here may
have implications for other physical problems, such as gas dynamics.

\subsection{The model}

Assume a long segment of a one lane road, deprived of entries and exits, only
moderately congested by traffic and free of breakdowns admitting a continuous
treatment, so as to permit us to postulate the usual equation of continuity:
\begin{equation}
\frac{\partial\rho}{\partial t} + u \frac{\partial\rho}{\partial x} =
- \rho\frac{\partial u}{\partial x}.
\label{cont}
\end{equation}
Here $\rho$ is the density of cars, the maximum of which $\rho_{\mathrm{max}}$
corresponds to a bumper to bumper
situation never occurring on our present model, and $u$ is the
local velocity. The right hand side of the second, Newtonian equation, formulated by
Payne \cite{Payne} and Whitham \cite{Whith1}, is less obvious:
\begin{equation}
\frac{\partial u}{\partial t} + u \frac{\partial u}{\partial x} =
\frac{V(\rho) - u}{\tau_0} - \frac{\nu_0}{\rho} \frac{\partial\rho}{\partial x} .
\label{mom}
\end{equation}
The first term on the right involves the mean drivers' reaction time $\tau_0$, and
the next term  models a diffusion effect depending on the drivers' awareness of
conditions beyond the preceding car. The constant $\nu_0$ is a parameter that
models the effect of gradients on the acceleration. The choice of $V(\rho)$
depends on the quality of the road.

An obvious solution is $\rho$ and $u = V(\rho)$ both constant. Whitham considers
this solution in his book and finds that is stable \cite{Whith1}.

In part I $V(\rho)$ was a linearly decreasing function of $\rho$:
\begin{equation}
V(\rho) = V_0 - h_0 \tau_0 \sqrt{\nu_0} \rho \equiv
V_0 (1 - \rho/\rho_{\mathrm{max}}), \qquad \rho \leq \rho_{\mathrm{max}},
\label{VrhoI}
\end{equation}
where the constant $h_0$ is related to $\rho_{\mathrm{max}}$  by
\begin{equation}
h_0 = \frac{V_0}{\rho_{\mathrm{max}} \tau_0 \sqrt{\nu_0}}.
\label{h0}
\end{equation}

In this paper the analysis will be restricted to $\rho \ll \rho_{\mathrm{max}}$
for which one can approximate $V(\rho)$ by $V_0$. In this connection we specify 
\begin{equation}
V(\rho) = V_0 = \mathrm{const}, \qquad \rho \leq \rho_{\mathrm{cr}}
\ll \rho_{\mathrm{max}}.
\label{Vrho}
\end{equation}
This is a form of $V(\rho)$ recently postulated for low density and high quality
of the road \cite{Belloq}--\cite{Dolfin}.

The model used in part I had most common sense properties. The flow of traffic
$Q = \rho V(\rho)$ increased from zero for zero density of cars, through a maximum
above which traffic becomes congested so that increasing the density no-longer
increases the flow, down to an extreme density preventing any motion. The flow
against density curve was a continuous parabola. All this is well enough. However,
this model can be improved on. When the distances between individual cars are long,
increase of density only results in a linear increase in the flow. If you cannot see
the cars preceding and following you, a possible small increase in the car density
will hardly be noticed. Thus the flow is a linear function of the density. Thus
the left hand portion of $V(\rho)$ should be a constant up to some
$\rho_{\mathrm{cr}}$ and the flow is $\rho V_0$. Calculations simplify as long as
we stay away from $\rho_{\mathrm{cr}}$. The value of $\rho_{\mathrm{cr}}$ will
depend on the quality of the road.

We introduce dimensionless variables by replacing
\begin{equation}
\fl t\ \to \ t\, \tau_0, \qquad (u,V_0) \to \ (u,V_0)\, \sqrt{\nu_0}, \qquad x \ \to
\ x\, \sqrt{\nu_0}\tau_0, \qquad \rho \to \ \rho\, (h_0\tau_0)^{-1}.
\label{reps}
\end{equation}
This leaves the continuity equation unchanged, and the Newtonian equation takes the
form:
\begin{equation}
\frac{\partial u}{\partial t} + u \frac{\partial u}{\partial x} =
V_0 - u - \frac{1}{\rho} \frac{\partial\rho}{\partial x} ,
\label{momds}
\end{equation}
slightly simpler than in part I. Here in part II we treat situations that never
become very congested ($\rho \ll \rho_{\mathrm{max}}$) which for the dimensionless
quantities requires that
\begin{equation}
\rho \leq \rho_{\mathrm{cr}} \ll V_0.
\label{aplc}
\end{equation}

Our model is a macroscopic one in which the traffic is treated as a fluid flow.
This is in contrast to the microscopic models involving motions of individual cars
and hybrid models combining elements of both. 

\section{Introducing Lagrangian coordinates}

(The Reader who has read part I can proceed to section 3.)

The non-linearity on the left hand side of equations (\ref{cont}) and (\ref{momds})
can be eliminated by introducing Lagrangian coordinates:
$\xi(x,t)$, the initial position (at $t=0$)
of a fluid element which at time $t$ was at $x$, and time $t$. In this description,
the independent variable $x$ becomes a function of $\xi$ and $t$, as are
the fluid parameters $\rho(\xi,t) = \rho(x(\xi,t),t)$ and
$u(\xi,t) = u(x(\xi,t),t)$.

Here and in what follows we adopt the convention that a superposition of two
functions which introduces a new variable is denoted by the same symbol as the
original function, but of the new variable. Denoting by $f$ either $\rho$ or $u$,
the basic transformation between Eulerian coordinates $x,t$ and Lagrangian ones
$\xi,t$ can be written as
\begin{equation}\label{trans}
\fl x(\xi,t) = \xi + \int_0^t u(\xi,t') \, \mathrm{d}t', \qquad
\frac{\partial x}{\partial t} = u(\xi,t), \qquad
\frac{\partial f(\xi,t)}{\partial t}
= \frac{\partial f(x,t)}{\partial t} + u \frac{\partial f(x,t)}{\partial x}.
\end{equation}
We denote by $s(\xi)$ the number of cars between the
last one at $\xi = \xi_{\mathrm{min}}$ and that at $\xi$:
\begin{equation}
s(\xi) = \int_{\xi_{\mathrm{min}}}^{\xi} \rho_0(\xi') \, \mathrm{d}\xi',
\qquad \rho_0(\xi) = \rho(x=\xi,t=0)
\label{sfxi}
\end{equation}
where $\rho_0(\xi)$ is the initial mass density distribution.

Here and in what follows, the subscript $0$ always refers to $t=0$. We will also use
the superscript $0$ to refer to $\xi=0$.

Here $\rho_0(\xi)$ is positive.
Hence $s(\xi)$ is an increasing function starting at $s(\xi_{\mathrm{min}}) = 0$,
and one can introduce a uniquely defined inverse function $\xi(s)$. The initial
position of a fluid element can be specified by either $\xi$ or $s$.

If a small initial interval $\mathrm{d}\xi$ at $t=0$ becomes $\mathrm{d}x$
at time $t$, mass conservation requires:
\begin{equation}
\mathrm{d}s = \rho_0(\xi) \mathrm{d} \xi = \rho(x,t) \mathrm{d}x.
\label{mc}
\end{equation}
This leads to a mass conservation equation in Lagrangian variables:
\begin{equation}\label{cont1}
\frac{\partial x(s,t)}{\partial s} = \frac{1}{\rho(s,t)},
\end{equation}
and to a useful operator identity
\begin{equation}\label{ss}
\frac{1}{\rho(x,t)}\frac{\partial}{\partial x} =
\frac{1}{\rho_0(\xi)}\frac{\partial}{\partial \xi} =
\frac{\partial}{\partial s}.
\end{equation}

Integrating (\ref{cont1}) over $s'$ from $s(\xi=0)$ to $s$, we obtain the continuity
equation in integral form:
\begin{equation}
X(s,t) \equiv x(s,t) - x(s^0,t) = \int_{s^0}^s \frac{\mathrm{d}s'}{\rho(s',t)},
\qquad s^0 = s(\xi=0).
\label{X}
\end{equation}
This indicates that if we know the car density in Lagrangian coordinates
$\rho(s,t)$, we can determine the evolving shape of the line of traffic, where
the distance $X$ is measured from the $\xi=0$ car.

The analogue of the continuity equation (\ref{cont}) is obtained by differentiating
(\ref{cont1}) by $t$. Using the middle part of (\ref{trans}) we obtain
\begin{equation}
\frac{\partial \psi(s,t)}{\partial t} = \frac{\partial u}{\partial s}, \qquad
\psi = \frac{1}{\rho}.
\label{contlv}
\end{equation}

The Newtonian equation in  Lagrangian coordinates is obtained from (\ref{momds})
and (\ref{ss}):
\begin{equation}
\frac{\partial u(s,t)}{\partial t} + u = V_0 - \frac{\partial\rho}{\partial s}.
\label{mom1}
\end{equation}
Equation (\ref{mom1}) can be solved to express $u(s,t)$ in terms of
$\rho$. Again, using the middle part of (\ref{trans}) we can also calculate
$x(s,t)$:
\begin{eqnarray}
u(s,t) &= \mathrm{e}^{-t} \biggl[\int_0^t N(s,t') \mathrm{e}^{t'} \, \mathrm{d}t' +
u(s,0) \biggr],
\label{uft}\\
N(s,t) &= V_0 - \frac{\partial \rho}{\partial s},\label{Nst}\\
x(s,t) &= \xi(s) + \int_0^t u(s,t') \, \mathrm{d}t'\nonumber\\
&= \xi(s) + u(s,0) - u(s,t) + \int_0^t N(s,t') \, \mathrm{d}t',\label{xst}
\end{eqnarray}
where the function $u(s,0)$ will be determined later.

\section{Finding the fluid density}

Differentiating the Newtonian equation (\ref{mom1}) by $s$, and using continuity
(\ref{contlv}), we obtain one equation for $\psi$:
\begin{equation}
\frac{\partial^2 \psi}{\partial t^2} - \frac{\partial}{\partial s} \Bigl( 
\frac{1}{\psi^2} \frac{\partial \psi}{\partial s} \Bigr) + 
\frac{\partial \psi}{\partial t} = 0 .
\label{psieq}
\end{equation}
In part I, there was an extra $\partial(1/\psi)/\partial s$ term, causing the
equation to have a symmetry when $\psi \to 1/\psi$, $s \to t$. Every solution had
a formal twin. This symmetry is now lost, even  though equation (\ref{psieq})
is simpler. 

Equation (\ref{psieq}) can be factorized in two possible ways, I and II:
\begin{equation}
\mathrm{I:\ } \qquad\Bigl( \frac{\partial}{\partial t} + \frac{\partial}{\partial s}
\frac{1}{\psi} \Bigr) \Bigl( \frac{\partial \psi}{\partial t} -
\frac{1}{\psi} \frac{\partial \psi}{\partial s} + \psi \Bigr) = 0 ,
\label{fact1}
\end{equation}
and
\begin{equation}
\mathrm{II:} \qquad\Bigl( \frac{\partial}{\partial t} - \frac{\partial}{\partial s}
\frac{1}{\psi} \Bigr) \Bigl( \frac{\partial \psi}{\partial t} +
\frac{1}{\psi} \frac{\partial \psi}{\partial s} + \psi \Bigr) = 0 .
\label{fact2}
\end{equation}

We will find that the second factor in (\ref{fact1}) best yields solutions such that
$X \geq 0$, whereas that in (\ref{fact2}) rules $X<0$, where $X$ is always the
distance from the car that started at $x=0$. In both (\ref{fact1}) and
(\ref{fact2}), one term in the second factor is absent as compared to part I.

In what follows, we will find solutions for which the second factor in one of
equations (\ref{fact1}), (\ref{fact2}) vanishes. We follow motion from left to
right. Factorization also means that we can only introduce the initial value of
the density (or $\psi$). The initial velocity $u(s,t=0)$ will then follow except
for a universal constant. We will have more to say about this later on.

The nonlinearities in (\ref{fact1}) and (\ref{fact2}) (second factors) can be
eliminated if one transforms the variables $s,t$ to $\eta,t$ in a 
way similar to the Lagrangian transformation (\ref{trans}):
\begin{equation}\label{trans1}
\fl s(\eta,t) = \eta \mp \int_0^t  \frac{\mathrm{d}t'}{\psi(\eta,t')}, \qquad
\frac{\partial s}{\partial t} = \mp \frac{1}{\psi(\eta,t)}, \qquad
\frac{\partial \psi(\eta,t)}{\partial t} = \frac{\partial \psi(s,t)}{\partial t}
\mp \frac{1}{\psi} \frac{\partial \psi}{\partial s} \, .
\end{equation}
Solving the resulting linear equation
\[
\frac{\partial \psi(\eta,t)}{\partial t} = - \psi
\]
we obtain, in view of the fact that s and $\eta$ are identical at $t=0$,
\begin{equation}
\psi(\eta,t) = \mathrm{e}^{-t} \psi_0(\eta), \qquad
\psi_0(\eta) \equiv \psi(s=\eta,0).
\label{psia}
\end{equation}
For this $\psi(\eta,t)$ we have
\begin{equation}
\int_0^t \frac{\mathrm{d}t'}{\psi} = \frac{\mathrm{e}^t - 1}{\psi_0(\eta)},
\label{inta}
\end{equation}
and finally, back to $\rho = 1/\psi$ and using (\ref{trans1}),
\begin{equation}
\fl s = \eta \mp \rho_0(\eta)A(t), \qquad
\rho_0(\eta) = \rho_0(s=\eta), \qquad A(t) = \mathrm{e}^t - 1.
\label{sfeta}
\end{equation}
In this relation, defining $s$ in terms of $\eta$ and $t$, $\rho_0(\eta)$ is defined
by (\ref{sfxi}) but is expressed in terms of $s$, where one has to rename $s$ to
$\eta$. The same procedure applies to $\psi_0(\eta)$ given by (\ref{psia}).

Using (\ref{psia}), we can express $\rho$ in terms of $\eta$ and $t$:
\begin{equation}
\rho(\eta,t) = \mathrm{e}^{t} \rho_0(\eta),
\label{rhoa}
\end{equation}
which tends to $\rho_0(s)$ as $t \to 0$.

We are now in a position to determine the function $u(s,0)$ needed in equations
(\ref{uft}) and (\ref{xst}). Differentiate (\ref{mom1}) by $s$ and then subtract
both sides of (\ref{psieq}) from the result to obtain
\begin{equation}
\Bigl(\frac{\partial}{\partial t} + 1\Bigr)\Bigl(\frac{\partial\psi}{\partial t} -
\frac{\partial u}{\partial s}\Bigr)= 0.
\end{equation}
Solved by
\begin{equation}
\frac{\partial\psi}{\partial t} - \frac{\partial u}{\partial s} = f(s)e^{-t}.
\end{equation}
Therefore, if $f(s)=0$, equation (\ref{contlv}) will be valid for all time.
All we require is 
\begin{equation}
\Bigl[\frac{\partial\psi}{\partial t} - \frac{\partial u}{\partial s} 
\Bigr]_{t=0} = 0, \qquad \mathrm{i.e.} \qquad \frac{\partial u(s,0)}{\partial s} =
\frac{\partial \psi_0}{\partial t}.
\end{equation}
This result, along with either (\ref{fact1}) or (\ref{fact2}), leads to
\begin{equation*}
\frac{\partial u(s,0)}{\partial s} = \pm \frac{1}{\psi_0}
\frac{\partial \psi_0}{\partial s} - \psi_0.
\end{equation*}
Integrating over $s'$ from $s^0\ (= s(\xi=0))$ to $s$ and transforming the result
to $\xi$, we end up with
\begin{equation}
\fl u(\xi,0) = u_0 - \xi \mp \ln \frac{\rho_0(\xi)}{a}, \qquad a = \rho_0(\xi=0),
\label{us0}
\end{equation}
where $u_0 = u(\xi=0,0) \geq 0$ is arbitrary.

The last task is to determine $u(s,t)$, $x(s,t)$, and $X(s,t)$, given by
(\ref{uft}), (\ref{xst}), and (\ref{X}), in terms of $\eta$. Using (\ref{rhoa}),
(\ref{sfeta}) and (\ref{Nst}) we find the integrand $N$: 
\begin{eqnarray}
N(s,t) &= V_0 - \frac{\partial \rho}{\partial s} =
V_0 - \frac{\partial \rho/\partial \eta}{\partial s/\partial \eta}\nonumber\\
&= V_0 \pm 1 - \frac{\pm 1 + \rho_0'(\eta)}{1 \mp
\rho_0'(\eta)(\mathrm{e}^t - 1)},\label{Nstpar} 
\end{eqnarray}
which tends to $V_0 \pm 1$ as $t \to \infty$.
Here $\eta = \eta(s,t)$ must be found as a solution of equation (\ref{sfeta})
which is often transcendental. If that is the case, the integrals (\ref{uft})
and (\ref{xst}) must be calculated numerically. On the other hand,
the integral in (\ref{X}) can be calculated analytically:
\begin{eqnarray}
X(s,t) &\equiv x(s,t) - x(s^0,t) = \int_{\eta^0}^{\eta}
\frac{\partial s'/\partial \eta'}{\rho(\eta',t)} \, \mathrm{d} \eta'
\nonumber\\
&= \mathrm{e}^{-t} \biggl[ \xi(s=\eta) - \xi(s=\eta^0)
\mp A(t) \ln \frac{\rho_0(\eta)}{\rho_0(\eta^0)}
\biggr],
\label{Xpar}
\end{eqnarray}
where $\eta = \eta(s,t)$ and $\eta^0 = \eta(s^0,t)$ are defined implicitly by
(\ref{sfeta}).

Restrictions on $\rho_0(\xi)$ are given in part I.

\section{Two exponential profiles of the initial fluid density}

We will see that two exponential profiles of the initial fluid density
\begin{eqnarray}
\rho_0(\xi) = a \exp (-\lambda \xi), \qquad& \xi \geq 0, \qquad \mathrm{i.e.}
\qquad \xi_{\mathrm{min}}=0,\label{ic}\\
%
%
\rho_0(\xi) = a \exp (\lambda \xi), \qquad& \xi \leq 0, \qquad \mathrm{i.e.}
\qquad \xi_{\mathrm{min}} = - \infty,
\label{ic2}
\end{eqnarray}
play a special role here, as in their case it is possible to eliminate the auxiliary
variable $\eta$, and even find the fluid density $\rho$ in terms of $X$ and
$t$.

Using equation (\ref{sfxi}) we first find
\begin{equation}
s^0 = s(\xi=0) = \int_{\xi_{\mathrm{min}}}^0 \rho_0(\xi') \, \mathrm{d}\xi' =
\cases{
0  &for (\ref{ic}),\\
\frac{a}{\lambda}  &for (\ref{ic2}),}
\label{s0}
\end{equation}
and then calculate
\begin{eqnarray}
s(\xi) &= s^0 + \int_0^{\xi} \rho_0(\xi') \, \mathrm{d}\xi' =
s^0 \mp \frac{a}{\lambda} \Bigl( \exp (\mp \lambda \xi) - 1 \Bigr)\nonumber\\
&=\cases{
\frac{a}{\lambda} \Bigl( 1 - \exp (-\lambda \xi) \Bigr)
& for (\ref{ic}),\\
\frac{a}{\lambda} \exp (\lambda \xi) &for (\ref{ic2}).}
\label{s2c}
\end{eqnarray}
The inverse functions are given by
\begin{equation}
\xi(s) = \mp \frac{1}{\lambda} \ln \Bigl( 1 \mp \case{\lambda}{a} (s - s^0) \Bigr)
= \cases{
- \frac{1}{\lambda} \ln \Bigl( 1 - \case{\lambda s}{a} \Bigr)
&for (\ref{ic}),\\
\frac{1}{\lambda} \ln \frac{\lambda s}{a} &for
(\ref{ic2}).}
\label{xi2c}
\end{equation}
Using this formula we can transform the initial conditions (\ref{ic}) and
(\ref{ic2}) given above in $x,t$ to $s,t$:
\begin{equation}
\rho_0(s) = a \mp \lambda (s - s^0)
= \cases{
a - \lambda s &for (\ref{ic}),\\
\lambda s &for (\ref{ic2}).}
\label{rho02c}
\end{equation}
We now look for solutions to equations (\ref{fact1}) and (\ref{fact2}) that recreate
the above initial conditions as $t$ tends to zero.

Replacing $s$ by $\eta$ in (\ref{rho02c}) and using the $\rho_0(\eta)$ so obtained
in (\ref{sfeta}) and (\ref{rhoa}), we obtain
\begin{equation}
s = \eta + A(t)\, \Bigl[ \lambda(\eta - s^0) \mp a \Bigr], \qquad
A(t) = \mathrm{e}^t - 1,
\label{sfeta1}
\end{equation}
leading to
\begin{equation}
\eta(s,t) = \frac{s + A(t)(\lambda s^0 \pm a)}{1 + \lambda A(t)},
\label{etafs1}
\end{equation}
and
\begin{equation}
\rho(s,t) = \frac{\mathrm{e}^t \rho_0(s)}{1 + \lambda A(t)} \equiv
\frac{\rho_0(s)}{\lambda + (1 - \lambda)\mathrm{e}^{-t}},\label{rhost1}
\end{equation}
where $\rho_0(s)$ is given by (\ref{rho02c}). In the limit $t \to 0$, we obtain
$\rho(s,t) \to \rho_0(s)$.

Using (\ref{rhost1}) and (\ref{us0}) we can determine $N(\xi,t)$ and $u(\xi,0)$
needed in equations (\ref{uft})--(\ref{xst}):
\begin{eqnarray}
N(\xi,t) &= V_0 \pm 1 \pm \frac{\lambda - 1}{1 + \lambda A(t)}
\nonumber\\
&\to V_0 \pm 1 \qquad \mathrm{as} \ t \to \infty,\label{Nst1}\\
u(\xi,0) &= u_0 + (\lambda - 1) \xi.
\label{us01}
\end{eqnarray}
Calculating the integrals in (\ref{uft}) and (\ref{xst}), we find the fluid
velocity $u(\xi,t)$ and characteristics $x(\xi,t)$ parametrized by the initial
fluid element position $\xi$:
\begin{eqnarray}
u(\xi,t) &= V_0 \pm 1 + \mathrm{e}^{-t}\nonumber\\
&\quad\times \biggl[ u_0 + (\lambda - 1)\xi - V_0 \mp 1\pm
\frac{\lambda - 1}{\lambda} \ln \Bigl( \lambda \mathrm{e}^t + 1 - \lambda
\Bigr)\Bigr]
\nonumber\\
&\to V_0 \pm 1 \qquad \mathrm{as} \ t \to \infty,\label{ufxit}\\
x(\xi,t) &= \xi + (V_0 \pm 1)t + (1 - \mathrm{e}^{-t})
[u_0 + (\lambda - 1) \xi - V_0 \mp1]\nonumber\\
&\quad\mp \frac{\lambda - 1}{\lambda} \biggl\{\biggl[
\mathrm{e}^{-t} + \frac{\lambda}{1 - \lambda}
\biggr]\ln [(1 - \lambda)\mathrm{e}^{-t} + \lambda] + t\mathrm{e}^{-t}
\biggr\}.
\label{xfxit}
\end{eqnarray}
We will now express $\rho$ directly in terms of $X$ by using (\ref{X}),
(\ref{rho02c}) and (\ref{rhost1}). The result is
\begin{equation}
\rho(X,t) = \frac{a}{\lambda + (1 - \lambda) \mathrm{e}^{-t}}
\exp\biggl(\mp\frac{\lambda}{\lambda + (1 - \lambda) \mathrm{e}^{-t}}X\biggr).
\label{rholim}
\end{equation}
The shapes evolve from $a \exp(\mp\lambda X)$ profiles to $(a/\lambda) \exp(\mp X)$
profiles as $t$ goes from $0$ to $\infty$. The relevant drawings, shown in figures
1 to 3, are very similar to figures 1 to 4 of part I. Total mass is conserved and
is $a/\lambda$ in each segment. This is now easily seen at all times (in part I
for $t \to \infty$ only).
\begin{figure}
\centerline{\includegraphics[scale=0.6]{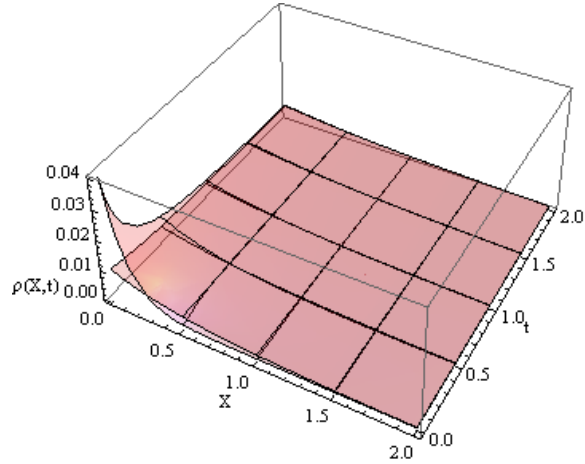}}
\caption{Two density profiles for various times as found
from our solution (I). Here $a =0.01$, $\lambda = 2$ in the first case, and
$a = 0.04$, $\lambda = 8$ in the second one. Nevertheless, the emerging  profiles
are seen to be identical after a while. The value of $a$ for each surface can be
seen as equal to $\rho(0,0)$.}
\end{figure}
\begin{figure}[b]
\centerline{\includegraphics[scale=0.6]{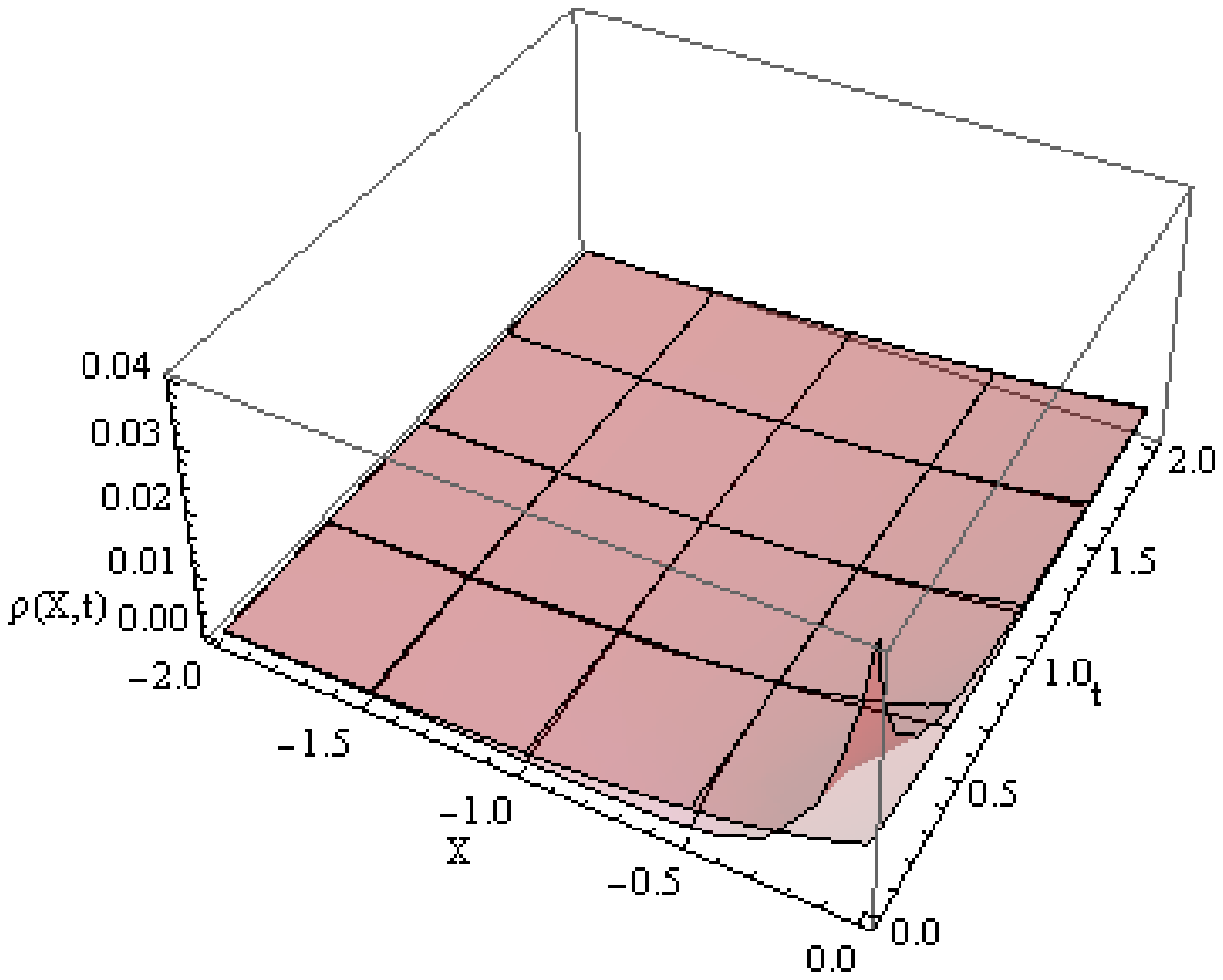}}
\caption{Two density profiles for various times as found
from our solution (II). Here $a =0.01$, $\lambda = 2$ in the first case, and
$a = 0.04$, $\lambda = 8$ in the second one. Nevertheless, the emerging  profiles
are seen to be identical after a while. The value of $a$ for each surface can be
seen as equal to $\rho(0,0)$.}
\end{figure}
\begin{figure}
\centerline{\includegraphics[scale=0.7]{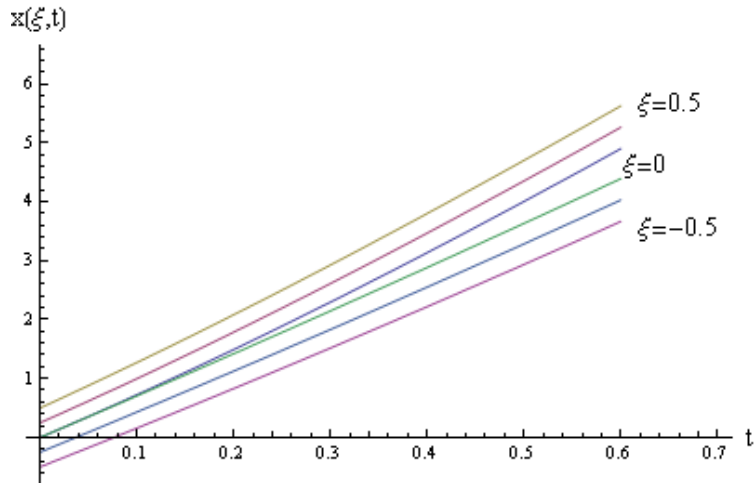}}
\caption{Characteristics $x(\xi,t)$ as functions of $t$
for $\xi=0,\: 0.25,\: 05,$ and $\xi=0,\: -0.25,\: -05$, $a=0.01$, $\lambda = 2$,
$u_0 = 7$, and $V_0 = 10$.}\label{charexp}
\end{figure}

\section{Initial density profiles that can be treated parametrically}

\begin{figure}
\centerline{\includegraphics[scale=0.6]{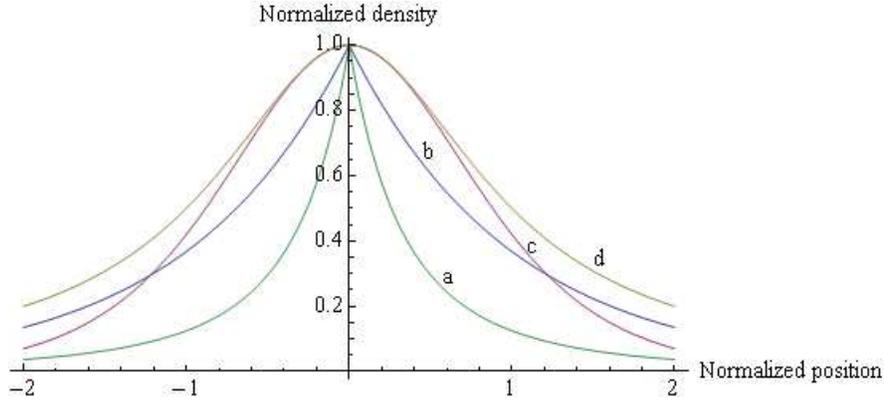}}
\caption{Normalized density profiles $\bar{\rho}_0=
\rho_0/a$ versus normalized position $\bar{\xi}=\lambda \xi$ for $\rho_0(\xi)$
given by (a): (\ref{ic}) and (\ref{ic2}), (b): (\ref{ic4}) for $b=1$,
$r=3$, (c): (\ref{icp}), and (d): (\ref{ic3}).}\label{densprof}
\end{figure}

In this section we present a few initial density profiles satisfying the
applicability conditions of our theory as formulated in part I,
see figure \ref{densprof}.

Detailed calculations will be performed for a pair of cases:
\begin{equation}
\rho_0(\xi) = \frac{a}{\cosh^2(\lambda \xi)} \equiv a \, \bigl[ 1 -
\tanh^2(\lambda \xi) \bigr],
\label{icp}
\end{equation}
where either $0 \leq \xi < \infty$ in case I, or $-\infty < \xi \leq 0$ in case II.

\begin{figure}
\centerline{\includegraphics[scale=0.6]{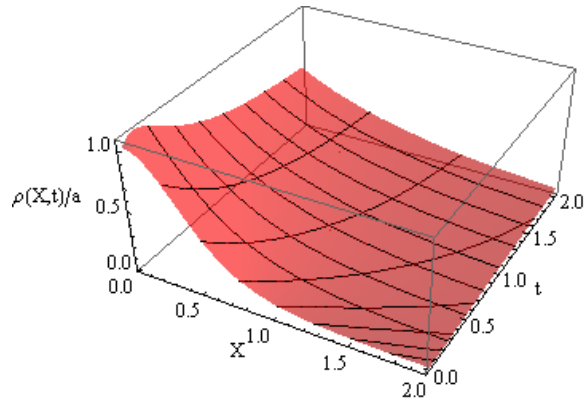}}
\caption{The normalized fluid density $\rho(X, t)/a$ represented parametrically
as found from our solution (I). Here $a =0.01$, $\lambda = 2$. The mesh lines
correspond to $t=\mathrm{const}$ or $\xi = 0,\:0.25,\:0.5,\:0.75, \dots $. Each
value of $\xi$ is equal to $X$ at $t=0$.}\label{densparI}
\end{figure}

The fact that the derivative $\mathrm{d}\rho_0(\xi)/\mathrm{d}\xi$ vanishes
at $\xi=0$, in contrast to the exponential profiles (\ref{ic}) and (\ref{ic2}), will
influence the time evolution in case I, see figure \ref{densparI}.

The remaining profiles will have a power law behaviour at infinity, $\rho_0(\xi)
\to (\pm\xi)^{-r}$ as $\pm\xi \to \infty$, where $r$ is a real number greater than
unity for integrability:
\begin{equation}
\rho_0(\xi) = \frac{a}{1 + (\lambda \xi)^2},
\label{ic3}
\end{equation}
and
\begin{equation}
\rho_0(\xi) = a \, \frac{b^r}{(\pm \lambda \xi + b)^r}, \qquad b > 0, \qquad r > 1,
\label{ic4}
\end{equation}
where the upper sign refers to case I, $\xi \geq 0$, and the lower one to case II,
$\xi \leq 0$.

By analogy to the exponential profiles (\ref{ic}) and (\ref{ic2}), each pair of
symmetric cases can be treated in a single calculation. For $\rho_0(\xi)$ given by
(\ref{icp}) we first find
\begin{equation}
s^0 = s(\xi=0) = \int_{\xi_{\mathrm{min}}}^0 \rho_0(\xi') \, \mathrm{d}\xi' =
\cases{
0 & for $\xi \geq 0$,\\
\frac{a}{\lambda} & for $\xi \leq 0$,}
\label{s0par}
\end{equation}
and then calculate
\begin{equation}
s(\xi) = s^0 + \int_0^{\xi} \rho_0(\xi') \, \mathrm{d}\xi' =
s^0 + \frac{a}{\lambda} \tanh(\lambda \xi).\label{sfxip}
\end{equation}
The inverse functions are given by
\begin{equation}
\xi(s) = \frac{1}{2\lambda} \ln
\frac{1 + \lambda (s - s^0)/a}{1 - \lambda (s - s^0)/a}
= \cases{
\frac{1}{2\lambda} \ln \frac{1 + \lambda s/a}{1 - \lambda s/a}
& for $\xi \geq 0$,\\
\frac{1}{2\lambda} \ln \frac{\lambda s/a}{2 - \lambda s/a}
& for $\xi \leq 0$.}
\label{xifsp}
\end{equation}
Using $\tanh(\lambda \xi)$ calculated from (\ref{sfxip}) in (\ref{icp}) we obtain
\begin{equation}
\rho_0(s) = a \Bigl[ 1 - \Bigl( \lambda (s - s^0)/a \Bigr)^2 \Bigr]
= \cases{
a [ 1 - (\lambda s/a)^2 ] & for $\xi \geq 0$,\\
\lambda s (2 - \lambda s/a) & for $\xi \leq 0$.}
\label{rhosp}
\end{equation}
Replacing here $s$ by $\eta$ and using the $\rho_0(\eta)$ so obtained in
(\ref{sfeta}) and (\ref{Nstpar}) along with (\ref{sfxip}) we find equations
defining $\eta(\xi,t)$ and the integrand $N(\eta,t)$ needed in equations
(\ref{uft})--(\ref{xst}):
\begin{eqnarray}
\frac{a}{\lambda} \tanh(\lambda \xi) = \eta - a [1 - (\lambda\eta/a)^2] A(t),
\qquad& \mathrm{for}\ \xi \geq 0,\label{etaeq1}\\
\frac{a}{\lambda} \bigl[ 1 + \tanh(\lambda \xi) \bigr] = 
\eta + \lambda \, \eta (2 - \lambda\eta/a) A(t),
\qquad& \mathrm{for} \ \xi \leq 0,\label{etaeq2}
\end{eqnarray}
\begin{equation}
N(\eta,t) = V_0 \pm 1 + \frac{\mp 1 - f(\eta)}{1 \mp f(\eta)A(t)},\label{Netat}
\end{equation}
where
\begin{equation}
f(\eta) =
\cases{
- 2 \eta \lambda^2/a + a & for $\xi \geq 0$,\\
2 \lambda \Bigl[ \eta (1 - \lambda/a)
+ 1 \Bigr] & for $\xi \leq 0$.}
\end{equation}

In a similar way we can determine $X(\eta,t)$ by using (\ref{Xpar}) along with
(\ref{xifsp}) and (\ref{rhosp}) with $s=\eta$:
\begin{eqnarray}
\fl X = - \frac{\mathrm{e}^{-t}}{\lambda} \biggl[\Bigl( \lambda A(t) + \frac{1}{2}
\Bigr) \ln
\frac{1 - \lambda\eta/a}{1 - \lambda\eta^0/a} + \Bigl( \lambda A(t) - \frac{1}{2}
\Bigr) \ln
\frac{1 + \lambda\eta/a}{1 + \lambda\eta^0/a}
\biggr] \qquad &\mathrm{for} \ \xi \geq 0, \label{X1p}\\
\fl X = \frac{\mathrm{e}^{-t}}{\lambda} \biggl[\Bigl( \lambda A(t) +
\frac{1}{2}\Bigr)
\ln\frac{\eta}{\eta^0} + \Bigl( \lambda A(t) - \frac{1}{2} \Bigr)
\ln \frac{2 - \lambda\eta/a}{2 - \lambda\eta^0/a}
\biggr] &\mathrm{for} \ \xi \leq 0. \label{X2p}
\end{eqnarray}
Surprisingly similar to the corresponding equation in part I. Again, just one term
has dropped out.

\begin{figure}[t]
\centerline{\includegraphics[scale=0.7]{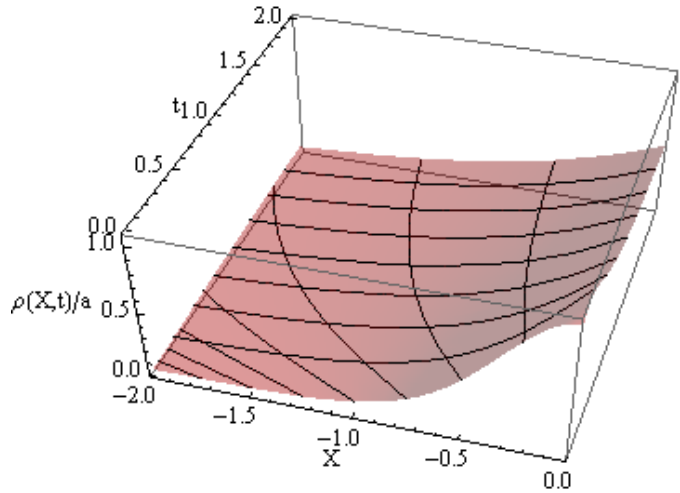}}
\caption{The normalized fluid density $\rho(X, t)/a$ as in figure 5 but as found
from our solution (II). Here again $a =0.01$, $\lambda = 2$, and $\xi = 0,\:-0.25,\:
-0.5,\:-0.75,\dots $, see $X$ at $t=0$.}\label{densparII}
\end{figure}

Using $X(\eta,t)$ given by these formulae and $\rho(\eta,t)$ given by (\ref{rhoa}),
where $\eta(\xi,t)$ is defined implicitly by either (\ref{etaeq1}) or
(\ref{etaeq2}), we obtain $\rho(X,t)$ in parametric form: $\rho(\xi,t)$ and
$X(\xi,t)$. This form is appropriate for the ParametricPlot3D command of
\textit{Mathematica}. The results are shown in figures \ref{densparI} and
\ref{densparII}. They resemble those shown in figures 6 and 7 of part I.

\begin{figure}
\centerline{\includegraphics[scale=0.7]{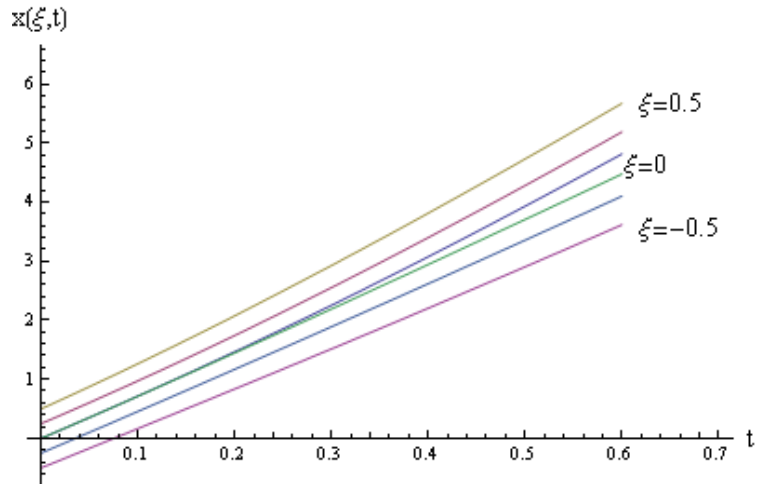}}
\caption{Characteristics $x(\xi,t)$ as functions of $t$
for $\xi=0,\: 0.25,\: 05,$ and $\xi=0,\: -0.25,\: -05$, $a=0.01$, $\lambda = 2$,
$u_0 = 7$, and $V_0 = 10$.}\label{charpar}
\end{figure}

The characteristics $x(\xi,t)$ can be found from equations (\ref{uft})--(\ref{xst})
by numerical integration, where the integrand $N(\xi,t)$ is defined by (\ref{Netat})
and either (\ref{etaeq1}) or (\ref{etaeq2}), and
\begin{equation}
u(\xi,0) = u_0 - \xi \mp \Bigl\{ \frac{a}{\lambda} \tanh(\lambda \xi) - 2 \,
\ln \Bigl[ \cosh(\lambda \xi) \Bigr] \Bigr\},\label{uxi0}
\end{equation}
see equations (\ref{us0}), (\ref{icp}) and (\ref{sfxip}). The results, depending
on two parameters $V_0$ and $u_0$, are shown in figure \ref{charpar}.

A characteristic feature of the plots representing the density given in parametric
form, $\rho(\xi,t)$ and $X(\xi,t)$, is that the mesh lines correspond to
$\xi = \mathrm{const}$, and $t = \mathrm{const}$, see figures \ref{densparI} and
\ref{densparII}. For the density
given explicitly, $\rho(X,t)$, they correspond to $X = \mathrm{const}$, and
$t = \mathrm{const}$. Each point on a $\xi = \mathrm{const}$ mesh line gives us
both the actual position $X$ and the associated density at time $t$,
for the car that started from $X = \xi$ at $t=0$. This information is given in the
frame moving with the discontinuity at $\xi=0$. The motion of these frames
in turn is described by the characteristics labeled $\xi=0$ in figure \ref{charpar}.

Adding cases I and II, we have a solution such that the initial configuration splits
in the middle, resulting once again in a slower cavalcade following a faster one,
see figure \ref{charpar}. This is rather like a two soliton solution of the
Korteweg--de Vries equation, see e.g. \cite{InfRol4}.

\section{Summary}

The LWRP model for traffic flow leaves the flow dependence on density open. This
dependence must be found for a specific road. Common sense implies some
ramifications. When there are no cars, flow is gone, so the $Q(\rho)$ curve
emerges from zero. A car a mile means no interaction, so $Q=V(\rho=0)\rho$ for a
while. As $\rho$ increases, interaction slows the growth of $Q(\rho)$ down until a
critical density is achieved. Now increase in density is balanced by the interaction
and $\mathrm{d}Q/\mathrm{d}\rho = 0$. Next $Q$ decreases down to zero at a complete
traffic jam density. Details vary from road to road, not to mention the make of
the cars. However, diagrams will have the following division in common:

\begin{enumerate}
 \item $\rho \leq \rho_{\mathrm{cr}}$, straight line indicating growth of $Q(\rho)$
 \item $Q(\rho)$ still grows, but at a decreasing pace, until a maximum is reached
 \item $Q(\rho)$ decreases with increasing $\rho$ down to zero at jam density.
\end{enumerate}

Here in part II we concentrated on the first region, whereas in part I
the whole curve was approximated by a parabola.
Differences were seen not be too important, especially for small
$a\  (= \rho_0(t = 0))$.

Our solutions both confirm and augment those of part I. They are somewhat similar
but simpler. Our exact solutions once again converge to single or double
stationary travelling wave structures after a few $\tau_0$, see figures 1 and 2.

It should be stressed that a complete solution is only possible if we combine our
two factorized equations, I and II.

The solutions presented here and in part I can be used to check numerical codes
before using them on more complicated situations. Simpler ones than here in part
II would be hard to find!

\end{document}